\documentclass[aps,preprint]{revtex4}%
\usepackage{amsfonts}
\usepackage{amsmath}
\usepackage{amssymb}
\usepackage{graphicx}%
\providecommand{\U}[1]{\protect\rule{.1in}{.1in}}

\begin{document}
\title[ ]{Scaling, Scattering, and Blackbody Radiation in Classical Physics}
\author{Timothy H. Boyer}
\affiliation{Department of Physics, City College of the City University of New York, New
York, New York 10031}

\begin{abstract}
Here we discuss blackbody radiation within the context of classical theory.
\ We note that nonrelativistic classical mechanics and relativistic classical
electrodynamics have contrasting scaling symmetries which influence the
scattering of radiation. \ Also, nonrelativistic mechanical systems can be
accurately combined with relativistic electromagnetic radiation only provided
the nonrelativistic mechanical systems are the low-velocity limits of fully
relativistic systems. \ Application of the no-interaction theorem for
relativistic systems limits the scattering mechanical systems for thermal
radiation to relativistic classical electrodynamic systems, which involve the
Coulomb potential. \ Whereas the naive use of nonrelativistic scatterers or
nonrelativistic classical statistical mechanics leads to the Rayleigh-Jeans
spectrum, the use of fully relativistic scatterers leads to the Planck
spectrum for blackbody radiation within classical physics. \ 

\end{abstract}
\maketitle

\subsection{Introduction}

The connections between classical and quantum physics are badly misunderstood
today. \ For example, the physics literature and the modern physics textbooks
claim that classical physics is incapable of accounting for the spectrum of
blackbody radiation and rather leads only to the divergent Rayleigh-Jeans
spectrum.\cite{texts} \ Actually, classical physics leads to the Planck
spectrum for blackbody radiation provided that one uses relativistic rather
than nonrelativistic classical physics and allows for classical zero-point
radiation, which is an intrinsic possibility of classical electrodynamics and
of thermodynamics.

Nonrelativistic classical mechanics was developed in the 17th and 18th
centuries, whereas classical electrodynamics was developed throughout the 19th
century into the early 20th. \ The years around the turn of the 20th century
saw the development of the theory of relativity. \ Electrodynamics was
relativistic whereas nonrelativistic mechanics did not satisfy the
requirements of special relativity. \ In the early 20th century, despite the
clear assessment that all of classical electrodynamics satisfies the ideas of
special relativity, physicists largely ignored the ideas of relativity in
connection with the unsolved problems of physics of that era. \ Current
textbooks of modern physics still teach the connections between classical and
quantum physics as though contemporary physicists were no better informed than
the physicists of the first half of the 20th century. \ Today textbooks of
modern physics often begin with a discussion of special relativity. \ However,
they fail to mention that the theory applies to all of relativistic classical
electrodynamics. \ Rather they claim,\cite{textrel} as was the viewpoint of
the physicists of the early 20th century, that special relativity needs to be
taken into account only in the physics of particles moving at a significant
fraction $v/c$ of the speed of light $c$. \ Thus when particles have high
velocity, one simply includes more terms in the ratio of $v/c$ for the
mechanical energy and momentum. \ However, this erroneous point of view
ignores the implications of the no-interaction theorem of Currie, Jordan and
Sudarshan\cite{No} which restricts the allowed interactions between
relativistic particles even when the particles are moving at small velocities. \ 

In the present article, we emphasize the contrast in scaling behavior between
nonrelativistic classical physics and relativistic classical electrodynamics.
\ We then show that this contrast in scaling behavior leads immediately to
contrasts in the scattering of electromagnetic radiation. \ The contrasts in
scattering behavior are tied directly to the ideas of thermal equilibrium for
blackbody radiation.

\subsection{Scaling Behavior within Classical Theory}

Any physical theory envisions a collection of elements which form the basis
for the theory. \ For example, nonrelativistic classical mechanics envisions a
set of masses $m_{i}$ at locations $\mathbf{r}_{i}$ moving with velocity
$\mathbf{v}_{i\text{ }}$which can interact through arbitrary potential
functions $V(|\mathbf{r}_{i}-\mathbf{r}_{j}|).$ In this theory, the masses,
lengths, and times all scale separately and continuously from 0 to $\infty.$
\ Thus the theory imagines the possibility of replacing any mass $m$ by a mass
$m^{\prime}=\sigma_{m}m,$ a length $l$ by a length $l^{\prime}=\sigma_{l}l,$
and a time $t$ by a time $t^{\prime}=\sigma_{t}t$ where $\sigma_{m}%
,~\sigma_{l},$ and $\sigma_{t}$ are arbitrary positive constants. \ Under such
a replacement, for example, a particle velocity $v=l/t$ becomes $v^{\prime
}=l^{\prime}/t^{\prime}=\sigma_{l}l/(\sigma_{t}t)=(\sigma_{l}/\sigma_{t})v.$
\ Kinetic energy $U=(1/2)mv^{2}$ is transformed as $U^{\prime}=(\sigma
_{m}\sigma_{l}^{2}/\sigma_{t}^{2})U.$ \ Thus in nonrelativistic mechanical
theory, any mechanical system can be replaced by a new mechanical system
which, for example, is twice as large, moves three times as fast, and has four
times the energy. \ 

Scaling in classical electrodynamics is quite different.\cite{FPscale}
\ Classical electrodynamics involves particles of charge $e$ and various
masses $m_{i}$ interacting through electromagnetic fields. \ Because of
certain universal constants found in nature, the relativistic classical
electrodynamic theory envisions elements which allow only a single
$\sigma_{ltU^{-1}}$-scaling, which maps length $l$ to $l^{\prime}%
=\sigma_{ltU^{-1}}l,$ time $t$ to $t^{\prime}=\sigma_{ltU^{-1}}t,$ and energy
$U$ to $U^{\prime}=U/\sigma_{ltU^{-1}}$ where $\sigma_{ltU^{-1}}$ is a
positive constant. \ The independent scalings of length $\sigma_{l}$ and of
time $\sigma_{t}$ envisioned within nonrelativistic mechanics are restricted
because of the existence of the universal constant $c$ (with dimensions of
$length/time$) corresponding to the speed of electromagnetic waves in vacuum.
\ The independent scalings of energy $\sigma_{U}$ ($\sigma_{U}=\sigma
_{m}\sigma_{l}^{2}/\sigma_{t}^{2}$) and of length $\sigma_{l}$ envisioned
within nonrelativistic mechanics are restricted because of the existence of
the universal constant $a_{S}/k_{B}^{4}$ (with dimensions of $1/(energy\times
length)^{3})$ corresponding to Stefan's constant $a_{S}$ divided by
Boltzmann's constant $k_{B}$ raised to the fourth power, and because of the
existence of a universal smallest electric charge $e$ (with dimensions of
$(energy\times length)^{1/2}).$ \ Thus any scaling of length $l$ and time $t$
where $l/t=c$ involves replacing $l/t$ by $l^{\prime}/t^{\prime}%
=\sigma_{ltU^{-1}}l/(\sigma_{ltU^{-1}}t)=l/t=c.$ \ Also any scaling of
potential energy $U$ and separation length $l$ where $U=e^{2}/l$ replaces $Ul$
by $U^{\prime}l^{\prime}=(U/\sigma_{ltU^{-1}})(\sigma_{ltU^{-1}}l)=Ul=e^{2}.$
\ It turns out that relativistic classical electrodynamics allows a complete
decoupling of the $\sigma_{ltU^{-1}}$-scale-invariant quantities (such as
angular momentum and velocity) from the quantities which are transformed by a
$\sigma_{ltU^{-1}}$-scale transformation (such as length, frequency, mass, and
energy). \ 

When discussing questions of scaling within relativistic classical
electrodynamics, one turns to angular momentum $J$ as the natural
$\sigma_{ltU^{-1}}$-scale-invariant parameter of choice. \ Of the three
familiar conserved quantities in the theory (energy $U$, linear momentum
$\mathbf{p}$, and angular momentum $\mathbf{J}$), only angular momentum is
$\sigma_{ltU^{-1}}$-scale invariant. \ Angular momentum is also an adiabatic
invariant.\cite{action} \ If one imagines electromagnetic radiation confined
to a spherical cavity with perfectly conducting walls, then a uniform
adiabatic compression (which changes the radius of the spherical cavity) will
leave the electromagnetic field angular momentum unchanged. \ Such an
adiabatic compression is often considered within thermodynamic analyses and so
is relevant for an understanding of blackbody radiation.

\subsection{Thermal Equilibrium for Electromagnetic Radiation}

In this article, we will consider only the \textit{classical} physics of
blackbody radiation. \ Blackbody radiation involves the radiation in a cavity
which has been brought to thermal equilibrium. \ The radiation in a cavity
with perfectly conducting walls will never achieve thermal equilibrium on its
own, since the scattering of electromagnetic radiation from a perfectly
reflecting wall may change the direction but not the frequency of the
radiation in the inertial frame at rest with respect to the walls of the
cavity. \ Rather, we must introduce radiation scatterers into the
mirror-walled cavity in order bring about thermal equilibrium. \ Crucially,
\textit{it is the character of the radiation scatterers which determines the
spectrum of radiation equilibrium.} \ 1) If nonrelativistic mechanical
scatterers are used,\cite{VV} or if nonrelativistic statistical mechanics is
applied to the scatterers or to the electromagnetic wave modes themselves,
then one arrives at the Rayleigh-Jeans spectrum.\cite{texts} \ This result is
noted in the textbooks of modern physics and throughout the physics
literature. \ 2) If relativistic electromagnetic scatterers are used and one
allows the natural possibility of classical electromagnetic zero-point
radiation, then one arrives at the Planck spectrum.\cite{BBrel} \ This result
is noted at a very few places in the physics literature and in none of the
textbooks. \ 3) If quantum mechanical scatterers are used, then one arrives at
the Planck spectrum. \ Indeed, the early rules of quantum physics were
developed as \textit{ad hoc} postulates introduced precisely in order to
obtain the Planck spectrum.\cite{Longair}

In the present article, we will carry out simple scattering calculations which
illustrate the contrasting aspects when nonrelativistic mechanical systems are
use as scatterers compared to when relativistic electromagnetic scatterers are
employed. \ In order to keep the analysis as simple and transparent as
possible, we consider the steady-state situation for a circularly-polarized
plane wave falling on a circular particle orbit for various particle masses
$m.$ \ Furthermore, again in the interests of simplicity, we will focus on the
scaling aspects of the scattering. \ 

\subsection{Radiation Scattering in Classical Physics}

\subsubsection{Scattering of a Plane Wave by a Nonrelativistic Mechanical
System}

We start with the traditional formulation of radiation scattering in classical
physics. \ The traditional treatment of the interaction of radiation and
matter in classical electrodynamics involves a nonrelativistic mechanical
system interacting with electromagnetic fields. \ Here we consider \ a
nonrelativistic particle of mass $m$ and charge $e$ moving in a circular orbit
of radius $r$ and frequency $\omega$ in a nonrelativistic central potential
$V(r).$ \ For simplicity, we take the potential as a power-law potential
$V(r)=\alpha r^{n+1}/(n+1),$ giving an attractive radial force of magnitude
$\mathbf{F}(r)=-\widehat{r}\alpha r^{n}$. \ Accordingly, Newton's second law
for the particle in the circular orbit takes the form%
\begin{equation}
m\omega^{2}r=\alpha r^{n}. \label{N2}%
\end{equation}
The mass $m$ and the constant $\alpha$ are associated with the fundamental
mechanical system itself while the frequency $\omega$ and the radius $r$
depend upon the energy or angular momentum contained within the system. \ We
will choose the angular momentum $J$ as the parameter for use in our analysis,
since angular momentum is a $\sigma_{ltU^{-1}}$-scale invariant and is also an
adiabatic invariant. \ In the nonrelativistic formulation, we have the angular
momentum $J$ for a circular orbit given by%

\begin{equation}
J=mr^{2}\omega.\label{Ang}%
\end{equation}
Solving equations (\ref{N2}) and (\ref{Ang}) for $r$ and $\omega$ in terms of
$m,~\alpha,$ and $J,$ we find%
\begin{equation}
r=\left(  \frac{J^{2}}{m\alpha}\right)  ^{1/(n+3)}\label{r1}%
\end{equation}
and
\begin{equation}
\omega=\left(  \frac{\alpha^{2}J^{n-1}}{m^{n+1}}\right)  ^{1/(n+3)}%
.\label{om1}%
\end{equation}
The velocity $v$ and energy $U$ of the orbiting charge are
\begin{equation}
v=r\omega=\frac{\alpha^{1/(n+3)}J^{(n+1)/(n+3)}}{m^{(n+2)/(n+3)}}\text{ \ and
\ }U=\frac{1}{2}mv^{2}+\frac{\alpha r^{n+1}}{n+1}=\frac{n+3}{2n+2}\frac
{\alpha^{2/(n+3)}J^{(2n+2)/(n+3)}}{m^{(n+1)/(n+3)}}.\label{vU}%
\end{equation}

A circularly-polarized electromagnetic plane wave of minimum amplitude $E_{0}$
falls on the mechanical system. \ If the circular orbit of the mechanical
system is in the $xy$-plane and the circular orbit is centered on the origin,
then we may take the plane wave as traveling along the $z$-axis\cite{circ}%
\cite{gaussian}%
\begin{align}
\mathbf{E}(z,t) &  =\widehat{i}E_{0}\cos(kz-\omega t)+\widehat{j}E_{0}%
\sin(kz-\omega t),\text{ \ \ \ }\nonumber\\
\text{\ }\mathbf{B}(z,t) &  =\widehat{j}E_{0}\cos(kz-\omega t)-\widehat{i}%
E_{0}\sin(kz-\omega t).\label{E-B}%
\end{align}
For minimum amplitude $E_{0}$ of the incident wave, the motion of the orbiting
charge must be in phase with the plane wave, with the direction of the
electric field in the direction of the particle velocity. \ Assuming a
steady-state situation, the power delivered to the orbiting charged particle
is $eE_{0}v=eE_{0}r\omega,$ and this must balance the power radiated by the
orbiting charge $P=(2/3)(e^{2}/c^{3})\omega^{4}r^{2}.$ \ Therefore we have%
\begin{equation}
eE_{0}r\omega=\frac{2e^{2}}{3c^{3}}\omega^{4}r^{2},\text{ \ \ \ \ or
\ \ \ }E_{0}=\frac{2e}{3c^{3}}\omega^{3}r,\label{E0}%
\end{equation}
where $r$ and $\omega$ are given in Eqs. (\ref{r1}) and (\ref{om1}), so that%
\[
E_{0}=\frac{2e}{3c^{3}}\frac{\alpha^{5/(n+3)}J^{(3n-1)/(n+3)}}%
{m^{(3n+4)/(n+3)}}.
\]
\ The magnetic field $\mathbf{B}$ of the plane wave places a $z$-component of
force on the orbiting charge; however, we will imagine the orbiting charge as
confined to the $xy$-plane by a frictionless surface and so will ignore this
force. \ 

\subsubsection{Scattering of Energy}

The plane wave in Eq. (\ref{E-B}) can be regarded as scattered by the charge
moving in a circular orbit. \ Energy and angular momentum are removed from the
incident wave and scattered into new directions. \ The total scattering
cross-section can be computed as the power $P$ radiated by the orbiting charge
divided by the power crossing per unit area $S=cE_{0}^{2}/(4\pi)$ in the plane
wave,%
\begin{align}
cross-section  &  =\frac{P}{S}=\left(  \frac{2e^{2}}{3c^{3}}\omega^{4}%
r^{2}\right)  \left(  \frac{4\pi}{cE_{0}^{2}}\right)  =\left(  \frac{2e^{2}%
}{3c^{3}}\omega^{4}r^{2}\right)  \frac{4\pi}{c}\left(  \frac{3c^{3}}%
{2e\omega^{3}r}\right)  ^{2}\nonumber\\
&  =\frac{6\pi c^{2}}{\omega^{2}}=6\pi c^{2}\left(  \frac{m^{n+1}}{\alpha
^{2}J^{n-1}}\right)  ^{\frac{2}{n+3}} \label{cs1}%
\end{align}
where we have used Eq. (\ref{E0}). \ 

In the low-velocity (nonrelativistic) limit, all the scattered radiation goes
into the fundamental mode which is at the same frequency $\omega$ as both the
particle orbital frequency and the circularly-polarized plane wave frequency.

The two most notable special cases are the simple harmonic oscillator
potential and the Coulomb potential. \ The simple harmonic potential
$V(r)=\alpha r^{2}/2$ involves $n=1,$ and $\alpha$ corresponds to the spring
constant. \ In this case, the scattering cross-section in Eq. (\ref{cs1})
becomes $cross-section=6\pi c^{2}(m/\alpha)=6\pi c^{2}/\omega_{0}^{2}$ which
depends upon the natural frequency of the oscillator $\omega_{0}%
=(\alpha/m)^{1/2}$ but is completely independent of the value of the
particle's angular momentum $J.$ \ The particle velocity $v$ is $v=\alpha
^{1/4}J^{1/2}/m^{3/4}$ and the particle energy is $U=(\alpha/m)^{1/2}%
J=\omega_{0}J.$ \ The electric field $E_{0}$ is related to the parameters of
the orbiting charge as $E_{0}=(2/3)(e/c^{3})(\alpha/m)^{5/4}(J/m)^{1/2}%
=(2/3)(e/c^{3})(\omega_{0})^{5/2}(J/m)^{1/2}.$ \ We see that, for fixed
angular momentum $J,$ different oscillators of the same natural frequency
$\omega_{0}=(\alpha/m)^{1/2}$ will have the same energy $U,$ but will involve
completely different values of velocity $v$ and electric field $E_{0}$
depending upon the choice of the mass $m.$

The Coulomb potential $V(r)=e^{2}/r$ involves $n=-2$ and $\alpha=e^{2}$. \ In
this case, we have $cross-section$ $=6\pi c^{2}[J^{3}/(me^{4})]^{2}%
=6\pi\lbrack e^{2}/(mc^{2})]^{2}(Jc/e^{2})^{6},$ where $[e^{2}/(mc^{2})]$ is
the classical radius of the electron, and $[e^{2}/(Jc)]$ is a $\sigma
_{ltU^{-1}}$-scale-invariant constant involving the angular momentum $J.$ The
particle velocity $v$ is $v=e^{2}/J$ and the particle energy is
$U=-(1/2)mc^{2}[e^{2}/(Jc)]^{2}.$ \ The amplitude of the circularly polarized
plane wave is $E_{0}=(2e/3)[mc^{2}/e^{2}]^{2}[e^{2}/(Jc)]^{7}.$ \ For the
Coulomb potential with fixed particle angular momentum, all quantities scale
with the mass $m.$ \ Thus the quantities $v,$ $U/m,$ $cross-section\times
m^{2}$ and $E_{0}/m^{2}$ depend upon only $\sigma_{ltU^{-1}}$-scale invariant
quantities. \ 

\subsection{Nonrelativistic Physics as a Limit of Relativistic Physics}

In the analysis just presented, we have used nonrelativistic mechanics for the
motion of the charged particle although we have used relativistic classical
electromagnetic theory for the circularly-polarized plane wave. \ Now we want
to check that the textbook claim that these scattering calculations are
justified in the sense that they represent low-velocity motion where $v/c<<1$
within a relativistic system. \ Thus we want to see that indeed our scattering
calculations represent low-velocity limits for a fully relativistic system.
\ Clearly the fully relativistic system must involve relativistic expressions
for the particle momentum $\mathbf{p}=m\gamma\mathbf{v}$ (where $\gamma
=(1-v^{2}/c^{2})^{-1/2})$ and angular momentum $\mathbf{J}=\mathbf{r\times p}$
(or in magnitude for a circular orbit $J=m\gamma vr).$ \ Then for a
relativistic circular orbit, the equations (\ref{N2}) and (\ref{Ang}) become%
\begin{equation}
m\gamma\omega^{2}r=\alpha r^{n}\label{N2R}%
\end{equation}
and
\begin{equation}
J=m\gamma r^{2}\omega\label{AngR}%
\end{equation}
where
\begin{equation}
\gamma=[1-(r\omega/c)^{2}]^{-1/2}.\label{gam}%
\end{equation}
However, despite the introduction of these relativistic expression for
momentum and angular momentum, our analysis is not relativistic. \ As
emphasized in the no-interaction theorem by Currie, Jordan, and
Sudarshan,\cite{No} any relativistic mechanical system which goes beyond point
interactions between particles must involve a field theory. \ The circular
orbit for our charged particle involves a potential energy $V(r)$ of
interaction between the particle of mass $m$ and a hypothetical particle of
very large mass $M\rightarrow\infty$ at the coordinate origin. \ In order to
become a relativistic theory, this potential energy function $V(r)$ must be
extended to a full relativistic field theory. \ In the case of the Coulomb
potential $V(r)=e^{2}/r$, the extension is thoroughly familiar as relativistic
classical electrodynamics. \ However, the relativistic field-theory extensions
of the other potentials, and in particular the harmonic oscillator potential,
are lacking. \ It is not sufficient to take the low-velocity limits of
relativistic expressions for particle energy and momentum; we must also deal
with interactions between particles which allow extension to a relativistic
field theory. \ 

\subsection{Radiation Scattering in Relativistic Classical Electrodynamics}

\subsubsection{Special Aspects of Scattering from a Charge in a Coulomb
Potential}

Having emphasized that the only familiar relativistic analysis involves the
use of relativistic classical electrodynamics, we now return to the scattering
calculation from a fully relativistic point of view. \ This means using fully
relativistic mechanical expressions and also limiting ourselves to the Coulomb
potential which is part of a fully relativistic field theory. \ We find that
there are now strong constraints on the behavior of the system. \ In this
electromagnetic case, the relativistic Eq. (\ref{N2R}) becomes%
\begin{equation}
m\gamma\omega^{2}r=e^{2}/r^{2},\label{N2RC}%
\end{equation}
while the angular momentum is still given by Eq. (\ref{AngR}). \ We can solve
Eqs. (\ref{AngR}) and (\ref{N2RC}) to obtain $r$ and $\omega$ as functions of
$m,$ $e^{2},$ and $J,$ giving%
\begin{equation}
r=\frac{e^{2}}{mc^{2}}\left(  \frac{Jc}{e^{2}}\right)  ^{2}\left[  1-\left(
\frac{e^{2}}{Jc}\right)  ^{2}\right]  ^{1/2}\text{ \ \ and \ \ }\omega
=\frac{mc^{3}}{e^{2}}\left(  \frac{e^{2}}{Jc}\right)  ^{3}\left[  1-\left(
\frac{e^{2}}{Jc}\right)  ^{2}\right]  ^{-1/2}.\label{romR}%
\end{equation}
The energy of the particle is
\[
U=mc^{2}-\frac{e^{2}}{r}=mc^{2}\left[  1-\left(  \frac{e^{2}}{Jc}\right)
^{2}\right]  ^{1/2}%
\]
and the velocity $v=r\omega$ of the particle in its circular orbit is%
\begin{equation}
v=r\omega=e^{2}/J.\label{vC}%
\end{equation}
The particle velocity is\ completely independent of the particle mass $m$.
\ Indeed, only the Coulomb potential has the orbiting particle velocity $v$
independent of the mass $m.$ \ This independence from $m$ is consistent with
the $\sigma_{ltU^{-1}}$-scale invariance of the velocity $v$ in relativistic
classical electrodynamics. \ Since the constants $e^{2}$ and $c,$ and also the
parameter $J$ are all $\sigma_{ltU^{-1}}$-scale invariant, the constant
$e^{2}/(Jc)$ is $\sigma_{ltU^{-1}}$-scale invariant. \ Thus the only parameter
giving a scale to the particle orbit is the mass $m$ which gives a length
scale corresponding to the classical radius of the electron $e^{2}/(mc^{2}).$
\ The mass $m$ also gives the scale of time in terms of the classical radius
of the electron divided by the speed of light $e^{2}/(mc^{3}),$ and the
frequency scale depends upon the inverse of this time. \ The electric field of
the incident plane wave also has its scale determined by the particle mass
$m.$ \ The power radiated in the relativistic treatment of the circular
orbit\cite{Jackson667} is $P=(2/3)(e^{2}/c^{3})\gamma^{4}\omega^{4}r^{2}$ so
that the relativistic version of Eq. (\ref{E0}) is%
\begin{equation}
E_{0}=\frac{2e}{3c^{3}}\gamma^{4}\omega^{3}r=\frac{2e}{3}\left(  \frac{mc^{2}%
}{e^{2}}\right)  ^{2}\left(  \frac{e^{2}}{Jc}\right)  ^{7}\left[  1-\left(
\frac{e^{2}}{Jc}\right)  ^{2}\right]  ^{-3}.\label{E0C}%
\end{equation}
Again this electric field magnitude is consistent with $\sigma_{ltU^{-1}}%
$-scaling since for a point charge the electric field behaves as
$\mathbf{E=}\widehat{r}e/r^{2},$ and $E_{0}$ in Eq. (\ref{E0C}) involves the
inverse of the classical radius of the electron squared times $\sigma
_{ltU^{-1}}$-scale-invariant quantities.

In the analysis involving a nonrelativistic low-velocity limit for the
orbiting charged particle, all the scattered radiation went into the radiation
at the same fundamental frequency $\omega$ as the orbital motion and the
incident plane wave. \ Such scattering cannot lead to thermal equilibrium
since there is no exchange of energy involving different frequencies.
\ However, in the fully relativistic calculation which is not restricted to
the limit of low velocities, the orbiting charged particle indeed shifts the
radiation of the incident wave into new frequencies; the scattering of the
plane wave moves part of the radiation to multiples of the fundamental
frequency and so acts like the sort of scatterer envisioned in discussions of
radiation thermal equilibrium. \ For a charged particle moving in a circular
orbit of radius $r$ at frequency $\omega,~$the power radiated per unit solid
angle into the $k^{th}$ harmonic is given by\cite{Jackson702}%
\begin{equation}
\frac{dP_{k}}{d\Omega}=\frac{e^{2}\omega^{4}r^{2}}{2\pi c^{3}}k^{2}\left\{
\left[  \frac{dJ_{k}(k\beta\sin\theta)}{d(k\beta\sin\theta)}\right]
^{2}+\left[  \frac{\cot\theta}{\beta}J_{k}(k\beta\sin\theta)\right]
^{2}\right\}  \label{dPdOm}%
\end{equation}
where here $\beta=r\omega/c,$ and $J_{k}$ is the Bessel function of order $k.$
\ The \textit{ratio} of the power radiated into the various harmonics depends
on only the velocity $\beta=v/c=e^{2}/(Jc)$ and is completely independent of
the mass $m$ of the charge in the circular orbit. \ This is the sort of
behavior which makes equilibrium thermal radiation independent of the details
of the relativistic electromagnetic scatterer. \ 

\subsubsection{Stability of the Orbiting Charge}

The total electromagnetic fields $\mathbf{E(r},t)=-\nabla\Phi-c^{-1}%
\partial\mathbf{A}/\partial t$ and$~\mathbf{B}(\mathbf{r},t)=\nabla
\times\mathbf{A}$ in space involve a homogeneous (source-free) solution of
Maxwell's equations (here given by the circularly-polarized plane wave) plus
the fields arising from the orbiting charge using the retarded Green function
for the scalar wave equation. \ In terms of the vector potential $\mathbf{A}$,
these expressions take the general form\cite{Jackson245}%
\begin{equation}
\mathbf{A(r},t)=\mathbf{A}^{in}(\mathbf{r},t)+\int\int d^{3}r^{\prime
}dt^{\prime}\frac{\delta(t-t^{\prime}-|\mathbf{r-r}^{\prime}|/c)}%
{|\mathbf{r-r}^{\prime}|}\frac{\mathbf{J(r}^{\prime},t^{\prime})}{c}\label{A}%
\end{equation}
where here $\mathbf{A}^{in}(\mathbf{r},t)$ corresponds to the incident
circularly-polarized plane wave and $\mathbf{J(r}^{\prime},t^{\prime})$
corresponds to the current source arising from the orbiting charge. \ In our
example involving the circularly-polarized plane wave, what prevents the
collapse of the orbiting charge into the center of the Coulomb potential is
the driving force of the homogeneous in-fields. \ The charged particle looses
energy and angular momentum due to radiation, but picks up energy and angular
momentum from the homogeneous (source-free) radiation fields, which in our
example are those of the circularly-polarized plane wave.

\subsubsection{Circular Particle Orbit in a Spherical Cavity}

In the discussion above, we considered the scattering of a
circularly-polarized plane wave by a charged particle in a circular orbit in a
Coulomb potential. \ The use of a circularly-polarized plane wave was made for
considerations of simplicity and familiarity. \ However, a still more relevant
calculation would involve a charged particle in steady-state motion in a
circular orbit in a Coulomb potential inside a spherical cavity with perfectly
conducting walls. \ Any accelerating electric charge would radiate so as to
introduce electromagnetic fields into the cavity. \ Thus steady-state motion
requires the presence of precisely those electromagnetic fields which meet the
boundary conditions at both the conducting walls and at the position of the
orbiting charged particle. \ The electromagnetic fields can be obtained by
modifying the calculations of Burko for a charged particle in a circular orbit
in free space.\cite{Burko} \ For a charge of mass $m$ in steady-state circular
orbit in a Coulomb potential at angular momentum $J$, the radius $R$ of the
cavity corresponds to certain discrete values related to the wavelength
$\lambda=2\pi c/\omega.$ There is radiation at all the harmonics of the
fundamental frequency, and the \textit{ratios} of the radiation energies at
the harmonics are independent of the fundamental frequency $\omega.$ \ If the
charged particle is replaced by a charged particle of the same charge and
different mass in the ratio $m^{\prime}/m$, the spherical cavity could be
compressed in the ratio $R^{\prime}/R=m/m^{\prime}$ so as to again bring the
orbital motion and radiation back into steady-state balance. \ The energy in
the radiation modes has the adiabatic invariant $U/\omega$ which is
$\sigma_{ltU^{-1}}$-scale invariant. \ 

\subsection{Classical Zero-Point Radiation and Blackbody Radiation}

\subsubsection{Classical Zero-Point Radiation}

In the scattering analysis used above, we have always worked with coherent
radiation connected to the steady-state motion of a charged particle.
\ However, thermal radiation involves not coherent but rather random radiation
over a spectrum of frequencies. \ As pointed out in an earlier
analysis\cite{SHO} of the thermodynamics of the harmonic oscillator (or of
radiation normal modes), the principles of thermodynamics allow the
possibility of zero-point energy. \ Zero-point energy is random energy which
exists even at the absolute zero of temperature and which takes the form
\begin{equation}
U_{0}(\omega)=const\times\omega\label{zpr}%
\end{equation}
for each normal mode of frequency $\omega.$ \ We note that this zero-point
energy satisfies $\sigma_{ltU^{-1}}$-scaling since energy $U$ and frequency
$\omega$ transform in the same fashion under the scaling. \ Since energy
divided by frequency $U/\omega$ is an adiabatic invariant for any oscillator
mode, the spectrum of zero-point radiation in a spherical cavity is invariant
under adiabatic compression provided that the constant $const$ is the same for
every radiation mode. \ Thus under adiabatic compression, a mode of frequency
$\omega$ becomes a mode of frequency $\omega^{\prime};$ however, the original
energy $U$ of the mode becomes energy $U^{\prime}$ of the new mode so that the
spectrum of the radiation in the compressed cavity is still that of zero-point
radiation, since it still takes the form $U_{0}^{\prime}(\omega^{\prime
})=const\times\omega^{\prime}.$

The motion of a relativistic charged particle in a Coulomb potential at the
center of the cavity is described by the action variables $J_{1},J_{2},J_{3}$
which have the units of angular momentum and are therefore $\sigma_{ltU^{-1}}%
$-scale invariant. \ The energy of the relativistic particle is given
by\cite{Goldstein}
\begin{equation}
U=m\gamma c^{2}-\frac{e^{2}}{r}=mc^{2}\left(  1+\frac{e^{2}}{[(J_{3}%
-J_{2})c+(J_{2}^{2}c^{2}-e^{2})^{1/2}]^{2}}\right)  ^{-1}\label{GJ}%
\end{equation}
which again satisfies $\sigma_{ltU^{-1}}$-scaling behavior with the particle
mass $m.$ \ The particle radiates away energy and angular momentum (just as
our charged particle in a steady-state circular orbit radiated away energy and
angular momentum into the radiation field), and picks up energy and angular
momentum out of the random radiation field (just as our orbiting particle
picked up energy and angular momentum from the circularly-polarized plane
wave). \ The constant $const$ appearing in the spectrum for the random
classical zero-point radiation is an adiabatic invariant and a $\sigma
_{ltU^{-1}}$-scale invariant; it takes the same value for each mode of the
radiation field. \ We expect that (just as in our example involving coherent
radiation above) the balance between energy pick up and loss for any charged
particle will lead to an average particle behavior which is related to that of
the random radiation. \ In this classical view, it is the random classical
zero-point radiation which prevents atomic collapse in a Coulomb potential.

\subsubsection{Classical Blackbody Radiation}

Random classical zero-point radiation is the unique spectrum of random
classical radiation which is $\sigma_{ltU^{-1}}$-scale invariant, Lorentz
invariant, invariant under adiabatic compression, and isotropic in every
inertial frame.\cite{any} \ Within classical electromagnetic theory,
zero-point radiation is crucial for understanding blackbody radiation. \ 

Based upon experimental measurements of Casimir forces,\cite{CasimirForces}
the constant $const$ appearing in the spectrum of classical zero-point
radiation takes the value%

\begin{equation}
const=\left(  \frac{\pi^{2}}{120c^{3}}\frac{k_{B}^{4}}{a_{S}}\right)
^{1/3}=0.527\times10^{-34}\mathbf{J\cdot s} \label{const}%
\end{equation}
where $a_{S}$ is Stefan's constant of 1789 and $k_{B}$ is Boltzmann's
constant. \ This constant takes the numerical value that is more familiarly
denoted as $\hbar/2$ where $2\pi\hbar=h$ is Planck's constant. \ This constant
value sets the scale for the energy $U_{0}(\omega)$ of radiation per normal
mode at every frequency $\omega$ at temperature $T=0.$ \ It will also set the
scale for the particle energy in Eq. (\ref{GJ}).

The classical understanding of thermal radiation is as follows. \ The
divergent spectrum of classical zero-point radiation is always present at any
temperature $T$. \ Thermal radiation at temperature $T>0$ represents a finite
density of radiation energy above the zero-point radiation. \ In a volume
$\mathcal{V},$ the total thermal energy equation of the blackbody spectrum
$U=a_{S}T^{4}\mathcal{V}$ (referring to to the energy above the zero-point
radiation) is invariant under $\sigma_{ltU^{-1}}$-scaling. \ Thus the quantity
$k_{B}T$ (like the quantity $U)$ has the dimensions of energy while the volume
$\mathcal{V}$ has the dimensions of length cubed, and the universal constant
$a_{S}/k_{B}^{4}$ is invariant under $\sigma_{ltU^{-1}}$-scaling. \ Therefore
the equation is $\sigma_{ltU^{-1}}$-scale invariant.

The presence of zero-point energy is contrary to the assumptions of
nonrelativistic classical statistical mechanics. \ But then too, the ideas of
special relativity are completely contrary to the ideas of nonrelativistic
classical statistical mechanics. \ It is nonrelativistic statistical mechanics
which suggests energy equipartition and the Rayleigh-Jeans spectrum. \ 

Classical zero-point radiation shares the $\sigma_{ltU^{-1}}$-scaling symmetry
of relativistic classical electrodynamics. \ Classical zero-point radiation
also provides the basis for understanding the Planck spectrum in classical
physics from a number of points of view. \ If we consider the thermodynamics
of a harmonic oscillator and ask for the smoothest interpolation between
zero-point energy $(1/2)\hbar\omega$ at low temperature and $k_{B}T$ at high
temperature, then we derive the Planck blackbody spectrum within classical
physics.\cite{SHO} \ If we compare paramagnetic behavior with diamagnetic
behavior in classical zero-point radiation while using relativistic limits
consistently, then we derive the Planck spectrum within classical
physics.\cite{dia} \ If we consider time-dilating conformal transformations of
thermal radiation in a Minkowski coordinate frame and in a Rindler frame, then
the Planck spectrum is derived within classical physics based upon the
structure of relativistic spacetime.\cite{BBrel}

\subsection{Discussion}

In our introductory courses on special relativity, we often discuss collisions
between relativistic particles. \ When we do this, we use only the mechanical
aspects of special relativity, only the expressions for mechanical energy
$U=m\gamma c^{2}$ and momentum $\mathbf{p}=m\gamma\mathbf{v}$. \ It is rarely
mentioned that if we attempt to introduce an interaction between the particles
which is not a point interaction, then relativity forces us to go to a full
field theory. \ This is the content of the no-interaction theorem of Currie,
Jordan, and Sudarshan.\cite{No} \ Mixtures of relativistic and nonrelativistic
physics follow neither the rules of relativistic nor of nonrelativistic
physics and are full of pitfalls for the unsuspecting physicist.\cite{contro}\ 

Electromagnetic radiation cannot bring itself to thermal equilibrium.
\ Therefore the analysis of blackbody radiation forces us to introduce
interactions between radiation and matter in order to describe thermal
equilibrium. \ However, if we hope to explain nature, we must use theories
which describe nature accurately. \ Use of nonrelativistic mechanics for the
scatterers of radiation leads to the same energy equipartition as appears in
nonrelativistic classical statistical mechanics. \ Accurate treatment of
thermal equilibrium within classical physics requires the use of a fully
relativistic analysis. \ Relativistic analysis leads to the Planck spectrum
for blackbody radiation within classical physics.


\begin{thebibliography}{99}
\bibitem {action}Indeed, all action variables (appearing in the action-angle
variables of mechanics) have the dimensions of angular momentum and are both
$\sigma_{ltU^{-1}}$-scale invariants and also adiabatic invariants.
\bibitem {gaussian}Since we are dealing with foundational questions of
electrodynamics, we will use gaussian units, which are natural units for the theory.
\end{thebibliography}
\end{document}